# Forced acoustical response of an open cavity coupled with a semi-infinite space

Yuhui Tong and Jie Pan[1]
School of Mechanical and Chemical Engineering,
The University of Western Australia,
Crawley, WA 6009, Australia

Yiwei Kou
Key Laboratory of Noise and Vibration Research, Institute of Acoustics,
Chinese Academy of Sciences,
100190, Beijing, China

[1] Author to whom correspondence should be addressed
Email: jie.pan@uwa.edu.au

**Abstract**

This paper presents a study of the forced acoustical response of an open cavity from the perspective of modal expansion. Based on the coupled mode theory, it is shown that the sound pressure distribution of an open cavity excited by a point source placed within the cavity can be expanded by a set of frequency-dependent eigenmodes, which are derived from the coupling between the cavity and a semi-infinite space. The calculation of the acoustical responses for baffled and unbaffled open cavities indicates that the proposed modal expansion converges with only a few frequency-dependent eigenmodes in the frequency range of interest. The results of this study eliminate the ambiguity involving the selection of appropriate basis functions, in modal expansion for the forced response problem in open cavities.

Keywords: open cavity; eigenmode; modal expansion.

PACS: 43.20.Fn, 43.20.Ks.

## I. INTRODUCTION

Open cavities are a class of acoustical systems in which the coupling between sound waves in a finite space and an infinite space plays a fundamental role in determining the system's modal characteristics and response. They are found in applications such as the wheel wells of aircraft,[1] parallel noise barriers,[2] and secondary volumes in large concert halls.[3] Although the properties of resonance modes in open cavities have been investigated extensively for passive and active noise control purposes,[4,5] difficulty remains when the system mode shapes (the spatial distribution of the sound pressure inside the cavity and outside in the open space) are used to expand the sound pressure response to a point source in the cavity. For example, when the radiated sound pressure by a point source within a pair of noise barriers was expanded by eigenmodes solved numerically using perfectly matched layers (PML), the correct response could only be found at the resonance frequencies.[6] The difficultymight be owing to the non-orthogonality and incompleteness of the *frequency-independent* eigenmodes (the resonance modes) associated with the acoustic resonance frequencies and modal damping ratio.[6–8] As a result, the sound pressure expanded by the frequency-independent eigenmodes of the open cavity may have a convergence issue, which inevitably gives rise to an ambiguity in the selection of basis functions, *i.e.*, if the resonance modes do not work for modal expansion, then what are the suitable basis functions?

Recent progress in modelling the sound scattering coefficients of open cavities due to an incident wave from a connected waveguide have demonstrated that the *frequency-dependent* eigensolutions of the effective Hamiltonian matrix of the sound field in the open cavity can be used to describe the coupling between the sound fields in the cavity and waveguides.[9–11] In this paper, the frequency-dependent eigenmodes is used to describe sound fields in the acoustically coupled open cavity and a semi-infinite space. The eigenvalue problem is developed at a given frequency for generating frequency-dependent eigenmodes and eigenvalues. Then the sound fields inside and outside the open cavity are expanded by these frequency-dependent eigenmodes. Numerical results are presented to confirm the accuracy of the proposed method and thus clarify the aforementioned ambiguity.

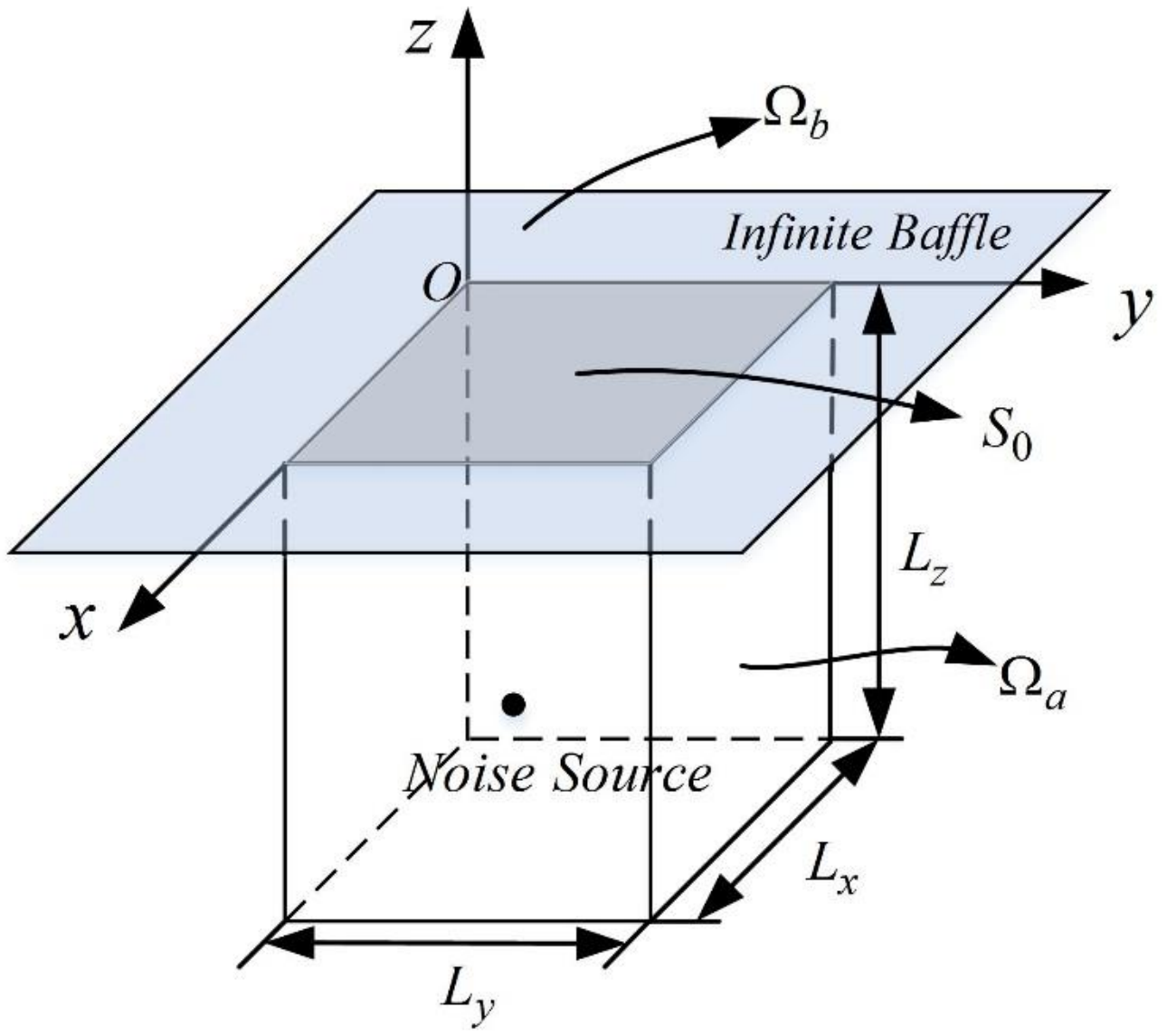

FIG. 1. A schematic diagram of a baffled open cavity.

## II. BAFFLED OPEN CAVITY

Figure 1 shows a model of a rectangular cavity with an opening located on an infinite baffle. For simplicity, all boundaries are assumed to be rigid. The acoustics such cavities have been studied for various purposes. For example, Wang *et al.*[12] adopted the traditional mode-matching technique to calculate the control field by a secondary source for the design of an active noise control system. Tam[13] investigated a similar two-dimensional (2D) configuration, via transformation methods, and determined the acoustic resonances of the open cavity. The modes Tam obtained are *frequency-independent*, of the same type studied in Refs. 4–8, which are not suitable for modal expansion. Meanwhile, it will be shown in the following part that the frequency-dependent eigenmodes are suitable basis functions for the correct modal expansion of the forced response of the open cavity.

### A. Formulation of solution (I): coupled mode theory

Omitting the time-dependence term, $e^{j\omega t}$, the sound pressure excited by a point source inside the cavity can be obtained by solving the nonhomogeneous Helmholtz equation:

$$\nabla^2 p + k^2 p = -q_s \delta(\boldsymbol{x} - \boldsymbol{x}_s), \tag{1}$$

where $k = \omega / c_0$ is the wavenumber, $c_0$ is the speed of sound, and $q_s$ and $\boldsymbol{x}_s$ are, respectively, the source strength and location.

Based on the coupled mode theory,[11,14,15] the space containing the sound pressure is first divided into finite and infinite sub-regions and the sound pressure in each sub-region is expanded by local basis functions. For a baffled open cavity, the whole space $\Omega$ is divided into the cavity $\Omega_a$ and the upper-half semi-infinite space $\Omega_b$, such that $p(\boldsymbol{x}) = p_i(\boldsymbol{x})$ for $\boldsymbol{x} \in \Omega_i$ (for $i$ = $a$ and $b$), *i.e.*:

$$p(\boldsymbol{x}) = \begin{pmatrix} p_a(\boldsymbol{x}) \\ p_b(\boldsymbol{x}) \end{pmatrix}. \tag{2}$$

In the cavity, the *closed*-cavity modes form a complete, orthogonal basis set, so that the sound pressure can be expanded as:

$$p_a(\boldsymbol{x}) = \sum_{\mu=0}^{\infty} \sum_{\nu=0}^{\infty} \sum_{\xi=0}^{\infty} a_{\mu,\nu,\xi} \phi_{\mu,\nu,\xi}(\boldsymbol{x}), \tag{3}$$

where $\phi_{\mu,\nu,\xi}(\boldsymbol{x}) = \psi_\mu(x)\psi_\nu(y)\psi_\xi(z)$ is the eigenmode of the *enclosed* rectangular cavity, $\psi_\mu(x) = \sqrt{\frac{2-\delta_{0,\mu}}{L_x}} \cos\left(\frac{\mu\pi}{L_x} x\right)$ and $\delta_{p.q}$ is the Kronecker delta function. In $\Omega_b$, however, the travelling wave solution would create complexity when plane waves are employed to expand the field in the upper-half space. Alternatively, in this paper, an expansion by an incomplete basis set is proposed, noting the fact that the sound source is located in $\Omega_a$ such that the sound field in $\Omega_b$ is dictated by the normal velocity distribution $v_\perp(\boldsymbol{x}')$, at the opening surface $S_0$ (the intersection between $\Omega_a$ and $\Omega_b$), *i.e.*:

$$p_b(\boldsymbol{x}) = jk\rho_0 c_0 \iint_{S_0} G_b(\boldsymbol{x}, \boldsymbol{x}') v_\perp(\boldsymbol{x}') dS', \tag{4}$$

where $G_b(\boldsymbol{x},\boldsymbol{x}') = \frac{1}{2\pi}\frac{e^{-jk|x-x'|}}{|x-x'|}$ is the Green's function for the upper half-space, $\rho_0$ is the air density, and the integral is evaluated over $S_0$. Since $\{\psi_m(x')\psi_n(y'), m, n = 0,1,2,\dots\}$ forms a complete basis set on $S_0$, and can be used for the expansion of the normal particle velocity on the surface, $v_\perp(\boldsymbol{x}')$ is rewritten as:

$$v_\perp(\boldsymbol{x}') = \sum_{m=0}^{\infty}\sum_{n=0}^{\infty} V_{m,n}\psi_m(x')\psi_n(y'). \tag{5}$$

Therefore, the contribution of $\psi_m(x')\psi_n(y')$ to the sound field is:

$$\varphi_{m,n}(\boldsymbol{x}) = jk\rho_0 c_0 \iint_{S_0} G_b(\boldsymbol{x},\boldsymbol{x}')\psi_m(x')\psi_n(y')dS', \tag{6}$$

which gives rise to:

$$p_b(\boldsymbol{x}) = \sum_{m=0}^{\infty}\sum_{n=0}^{\infty} V_{m,n}\varphi_{m,n}(\boldsymbol{x}), \tag{7}$$

despite the fact that $\{\varphi_{m,n}(\boldsymbol{x})\}$ does not form a complete basis set for all functions defined in $\Omega_b$.

Then, the fields in two sub-regions are coupled through continuity conditions at the intersection. In $\Omega_a$, $p_a$ satisfies:

$$\nabla^2 p_a(\boldsymbol{x}) + k^2 p_a(\boldsymbol{x}) = -q_s\delta(\boldsymbol{x}-\boldsymbol{x}_s), \tag{8}$$

together with corresponding boundary conditions, while $\phi_{\mu,\nu,\xi}(\boldsymbol{x})$ satisfies:

$$\nabla^2\phi_{\mu,\nu,\xi} + k_{\mu,\nu,\xi}^2\phi_{\mu,\nu,\xi} = 0, \tag{9}$$

$$k_{\mu,\nu,\xi}^2 = (\mu\pi/L_x)^2 + (\nu\pi/L_y)^2 + (\xi\pi/L_z)^2, \tag{10}$$

and rigid boundary conditions for all six walls of the rectangular cavity, including $S_0$. Multiplying Eq. (8) and Eq. (9) by $\phi_{\mu,\nu,\xi}$ and $p_a$ respectively and taking the difference of the resulting equations yields:

$$(p_a\nabla^2\phi_{\mu,\nu,\xi} - \phi_{\mu,\nu,\xi}\nabla^2 p_a) + (k_{\mu,\nu,\xi}^2 - k^2)p_a\phi_{\mu,\nu,\xi} = q_s\phi_{\mu,\nu,\xi}\delta(\boldsymbol{x}-\boldsymbol{x}_s). \tag{11}$$

Integrating Eq. (11) over $\Omega_a$, and applying Green's theorem gives rise to:

$$jk\rho_0 c_0 \iint_{S_0}\phi_{\mu,\nu,\xi}v_\perp dS_0 + (k_{\mu,\nu,\xi}^2 - k^2)a_{\mu,\nu,\xi} = q_0\phi_{\mu,\nu,\xi}(\boldsymbol{x}_s). \tag{12}$$

The above equation can be further simplified into:

$$\sum_{m=0}^{\infty}\sum_{n=0}^{\infty} jk\psi_\xi(0)\delta_{\mu,m}\delta_{\nu,n}\rho_0 c_0 V_{m,n} + (k_{\mu,\nu,\xi}^2 - k^2)a_{\mu,\nu,\xi} = q_0\phi_{\mu,\nu,\xi}(\boldsymbol{x}_s), \tag{13}$$

by using Eq. (5). Another constraint is the continuity condition for sound pressure at the opening, *i.e.*, $p_a|_{S_0} = p_b|_{S_0}$, such that:

$$\sum_{\mu'=0}^{\infty}\sum_{\nu'=0}^{\infty}\sum_{\xi'=0}^{\infty} a_{\mu',\nu',\xi'}\psi_{\mu'}(x)\psi_{\nu'}(y)\psi_{\xi'}(0) = \sum_{m=0}^{\infty}\sum_{n=0}^{\infty} V_{m,n}\varphi_{m,n}(\boldsymbol{x}). \tag{14}$$

Multiplying $\psi_\mu(x)\psi_\nu(y)$ and integrating over the interface leads to:

$$\sum_{\mu'=0}^{\infty}\sum_{\nu'=0}^{\infty}\sum_{\xi'=0}^{\infty} \delta_{\mu,\mu'}\delta_{\nu,\nu'}\psi_{\xi'}(0)a_{\mu',\nu',\xi'} = \rho_0 c_0\sum_{m=0}^{\infty}\sum_{n=0}^{\infty} Z_{\mu,\nu,m,n}V_{m,n}, \tag{15}$$

where $Z_{\mu,\nu,m,n}$ is the radiation impedance of a baffled rectangular surface[16] of size $L_x \times L_y$:

$$Z_{\mu,\nu,m,n} = jk \iint_{S_0} \iint_{S_0} \psi_\mu(x)\psi_\nu(y) \frac{e^{-jk\sqrt{(x-x')^2+(x-y')^2}}}{2\pi\sqrt{(x-x')^2+(x-y')^2}} \psi_m(x')\psi_n(y')dS'dS. \tag{16}$$

Using Eqs. (13) and (15), vectors $\boldsymbol{a} = [\cdots \quad a_{\mu,\nu,\xi} \quad \cdots]^T$ and $\boldsymbol{V} = [\cdots \quad V_{m,n} \quad \cdots]^T$ can be determined by solving:

$$\boldsymbol{HV} + (\boldsymbol{K} - k^2\boldsymbol{I})\boldsymbol{a} = \boldsymbol{S}, \tag{17}$$

and:

$$\boldsymbol{Ma} = \boldsymbol{ZV}, \tag{18}$$

where the corresponding matrices are defined as follows: $\boldsymbol{H}_{(\mu,\nu,\xi),(m,n)} = jk\delta_{\mu,m}\delta_{\nu,n}\psi_\xi(0)$, $\boldsymbol{K}_{(\mu,\nu,\xi),(\mu',\nu',\xi')} = k^2_{\mu,\nu,\xi}\delta_{\mu,\mu'}\delta_{\nu,\nu'}\delta_{\xi,\xi'}$ , $\boldsymbol{S} = q_s[\cdots \quad \phi_{\mu,\nu,\xi}(\boldsymbol{x}_s) \quad \cdots]^T$ , $\boldsymbol{M}_{(m,n),(\mu,\nu,\xi)} = \delta_{\mu,\mu'}\delta_{\nu,\nu'}\psi_{\xi'}(0)$, $\boldsymbol{Z}_{(\mu,\nu),(m,n)} = Z_{\mu,\nu,m,n}$. With the definition of these matrices, Eqs. (17) and (18) can be further reduced to:

$$(\boldsymbol{D} - \boldsymbol{k^2})\boldsymbol{a} = \boldsymbol{S}, \tag{19}$$

where $\boldsymbol{D} = \boldsymbol{K} - \boldsymbol{HZ}^{-1}\boldsymbol{M}$ is known as the effective Hamiltonian[17] (reduced differential operator) of the system. Solving $\boldsymbol{a}$ and $\boldsymbol{V}$ respectively from Eqs. (19) and (18) and substituting them into Eqs. (3) and (7) results in the desired sound pressure distribution.

**B. Formulation of solution (II): bi-orthogonal basis and modal expansion**

The homogeneous Eq. (19) forms the following eigenvalue problem (EVP):

$$\boldsymbol{D}\boldsymbol{g}_{\mu,\nu,\xi} = K^2_{\mu,\nu,\xi}\boldsymbol{g}_{\mu,\nu,\xi}, \tag{20}$$

where $K^2_{\mu,\nu,\xi}$ is the eigenvalue and the eigenvector $\boldsymbol{g}_{\mu,\nu,\xi}$ satisfies the bi-orthogonal relation:[10]

$$\boldsymbol{g}^T_{\mu',\nu',\xi'}\boldsymbol{g}_{\mu,\nu,\xi} = \delta_{\mu',\mu}\delta_{\nu,\nu'}\delta_{\xi,\xi'}\boldsymbol{g}^T_{\mu,\nu,\xi}\boldsymbol{g}_{\mu,\nu,\xi}. \tag{21}$$

An alternative expression of Eq. (21) is:

$$\iiint_{V_0} \Phi_{\mu',\nu',\xi'}(\boldsymbol{x})\Phi_{\mu,\nu,\xi}(\boldsymbol{x})dV = \delta_{\mu',\mu}\delta_{\nu,\nu'}\delta_{\xi,\xi'} \iiint_{V_0} \Phi^2_{\mu,\nu,\xi}(\boldsymbol{x})dV, \tag{22}$$

where $\Phi_{\mu,\nu,\xi}(\boldsymbol{x})$ is the eigenfunction corresponding to $\boldsymbol{g}_{\mu,\nu,\xi}$ such that:

$$\Phi_{\mu,\nu,\xi}(\boldsymbol{x}) = \boldsymbol{g}_{\mu,\nu,\xi}\boldsymbol{\phi}, \tag{23}$$

and $\boldsymbol{\phi} = [\cdots \quad \phi_{\mu',\nu',\xi'}(\boldsymbol{x}) \quad \cdots]$. It is worth noting that the EVP defined in Eq. (20) is (*source*) *frequency-dependent*, so the eigenvalues and eigenvectors/eigenfunctions depend on the source frequency $k$ as well. It then transpires that utilizing these frequency-dependent eigenmodes enables the expansion of the sound field of the open cavity. Expanding $\boldsymbol{a}$ into $\{\boldsymbol{g}_{\mu,\nu,\xi}\}$:

$$\boldsymbol{a} = \sum_{\mu'=0}^{\infty}\sum_{\nu'=0}^{\infty}\sum_{\xi'=0}^{\infty} c_{\mu',\nu',\xi'}\boldsymbol{g}_{\mu',\nu',\xi'}, \tag{24}$$

and making the substitution into Eq. (18) yields:

$$c_{\mu,\nu,\xi} = \frac{\boldsymbol{g}^T_{\mu,\nu,\xi}\boldsymbol{S}}{(K^2_{\mu,\nu,\xi} - k^2)\boldsymbol{g}^T_{\mu,\nu,\xi}\boldsymbol{g}_{\mu,\nu,\xi}}. \tag{25}$$

Combining Eq. (24) together with Eqs. (3), (7), and (18) leads to the expression of modal expansion:

$$p(\boldsymbol{x}) = \begin{pmatrix} p_a(\boldsymbol{x}) \\ p_b(\boldsymbol{x}) \end{pmatrix} = \sum_{\mu=0}^{\infty} \sum_{\nu=0}^{\infty} \sum_{\xi=0}^{\infty} c_{\mu,\nu,\xi} \begin{pmatrix} \Phi_{\mu,\nu,\xi}(\boldsymbol{x}) \\ \Psi_{\mu,\nu,\xi}(\boldsymbol{x}) \end{pmatrix}, \quad (26)$$

where $\Psi_{\mu,\nu,\xi}(\boldsymbol{x})$ is given by:

$$\Psi_{\mu,\nu,\xi}(\boldsymbol{x}) = \boldsymbol{\varphi}^T \boldsymbol{Z}^{-1} \boldsymbol{M} \boldsymbol{g}_{\mu,\nu,\xi} \quad (27)$$

and $\boldsymbol{\varphi} = [\cdots \quad \varphi_{m,n}(\boldsymbol{x}) \quad \cdots]$.

**C. Numerical validation**

A numerical investigation is conducted to examine the analytical solution obtained in Secs. II.A and II.B. The cavity in Fig. 1 has a configuration of 0.432 m long ($L_x$), 0.67 m wide ($L_y$), and 0.598 m high ($L_z$), the same as what was considered in Ref. 12. The source is located at (0.1, 0.1, $-L_z + 0.1$) m while the evaluation points inside and outside the cavity are randomly chosen at (0.2, 0.3, $-L_z + 0.4$) m and (1.3, 1.4, $-L_z + 1.5$) m. The analytical method proposed in Secs. II.A and II.B is obtained with MATLAB codes, when 140 *closed*-cavity modes $\{\phi_{\mu,\nu,\xi}(\boldsymbol{x})\}$ and seven external modes $\{\varphi_{m,n}(\boldsymbol{x})\}$ are used for local expansion and hence the computation of frequency-dependent eigenmodes $\begin{pmatrix} \Phi_{\mu,\nu,\xi}(\boldsymbol{x}) \\ \Psi_{\mu,\nu,\xi}(\boldsymbol{x}) \end{pmatrix}$.

TABLE I. The first fifteen modes of the closed and open rectangular cavities, and the corresponding frequencies; $f_{\mu,\nu,\xi} = k_{\mu,\nu,\xi} c_0/2\pi$ for the closed cavity, $F_{\mu,\nu,\xi} = K_{\mu,\nu,\xi} c_0/2\pi$ for the open cavity, at source frequency $f = 500$ Hz.

| $\mu$ | $\nu$ | $\xi$ | $f_{\mu,\nu,\xi}$ (Hz) | $F_{\mu,\nu,\xi}$ (Hz) |
|---|---|---|---|---|
| 0 | 0 | 0 | 0 | 133.26+23.55$j$ |
| 0 | 1 | 0 | 253.73 | 283.08+8.64$j$ |
| 0 | 0 | 1 | 284.28 | 382.88+71.04$j$ |
| 0 | 1 | 1 | 381.04 | 449.37+40.02$j$ |
| 1 | 0 | 0 | 393.51 | 413.08+3.68$j$ |
| 1 | 1 | 0 | 468.22 | 484.95+1.83$j$ |
| 1 | 0 | 1 | 485.46 | 544.96+20.79$j$ |
| 0 | 2 | 0 | 507.46 | 522.01+2.07$j$ |
| 1 | 1 | 1 | 547.77 | 603.52+11.06$j$ |
| 0 | 0 | 2 | 568.56 | 604.31+75.40$j$ |
| 0 | 2 | 1 | 581.66 | 631.25+11.13$j$ |
| 0 | 1 | 2 | 622.60 | 666.71+46.10$j$ |
| 1 | 2 | 0 | 642.16 | 655+0.40$j$ |
| 1 | 0 | 2 | 691.46 | 743.95+28.66$j$ |
| 1 | 2 | 1 | 702.03 | 752.1+3.30$j$ |

A total of 140 frequency-dependent eigenmodes are obtained via the analytical method for an open rectangular cavity. Table I lists the first fifteen eigensolutions when the source frequency is $f = 500$ Hz ($k = 9.24$). The corresponding natural frequencies of the closed cavity are also listed for comparison. It is clear that the eigenfrequency of each open-cavity mode becomes complex, in which the imaginary part is related to the radiation loss. Figure 2 plots slices of the modulus of the modal function of the $(\mu, \nu, \xi)$ mode, $|\Phi_{\mu,\nu,\xi}|$ ($\Psi_{\mu,\nu,\xi}(\boldsymbol{x})$ is not plotted for this case because of the high computational expense, as every point value evolves a Raleigh integral). The nodal lines are distinguishable for these low frequency eigenmodes, which justifies the inheritance of the closed-cavity modes' indexes $(\mu, \nu, \xi)$ to classify the open-cavity modes. The bi-orthogonality of the eigensolutions is validated and shown in Table II by calculating:

$$A_{(\mu,\nu,\xi),(\mu',\nu',\xi')} = |\boldsymbol{g}^T_{\mu,\nu,\xi}\boldsymbol{g}_{\mu,\nu,\xi}|, \tag{28}$$

where $\boldsymbol{g}_{\mu,\nu,\xi}$ is normalized such that $|\boldsymbol{g}^\dagger_{\mu,\nu,\xi}\boldsymbol{g}_{\mu,\nu,\xi}| = 1$.

TABLE II. Values of $A_{(\mu,\nu,\xi),(\mu',\nu',\xi')}$ for first ten eigenmodes, at source frequency $f = 500$ Hz. In the table, values below $10^{-13}$ are taken as 0.

| $(\mu,\nu,\xi)$ \ $(\mu',\nu',\xi')$ | (0,0,0) | (0,1,0) | (0,0,1) | (1,1,0) | (0,1,1) | (1,0,0) | (0,2,0) | (1,0,1) | (1,1,1) | (0,0,2) |
|---|---|---|---|---|---|---|---|---|---|---|
| (0,0,0) | 0.9517 | 0 | 0 | 0 | 0 | 0 | 0 | 0 | 0 | 0 |
| (0,1,0) | 0 | 0.9717 | 0 | 0 | 0 | 0 | 0 | 0 | 0 | 0 |
| (0,0,1) | 0 | 0 | 0.7803 | 0 | 0 | 0 | 0 | 0 | 0 | 0 |
| (1,1,0) | 0 | 0 | 0 | 0.9889 | 0 | 0 | 0 | 0 | 0 | 0 |
| (0,1,1) | 0 | 0 | 0 | 0 | 0.8956 | 0 | 0 | 0 | 0 | 0 |
| (1,0,0) | 0 | 0 | 0 | 0 | 0 | 0.9962 | 0 | 0 | 0 | 0 |
| (0,2,0) | 0 | 0 | 0 | 0 | 0 | 0 | 0.9890 | 0 | 0 | 0 |
| (1,0,1) | 0 | 0 | 0 | 0 | 0 | 0 | 0 | 0.9541 | 0 | 0 |
| (1,1,1) | 0 | 0 | 0 | 0 | 0 | 0 | 0 | 0 | 0.9835 | 0 |
| (0,0,2) | 0 | 0 | 0 | 0 | 0 | 0 | 0 | 0 | 0 | 0.7098 |

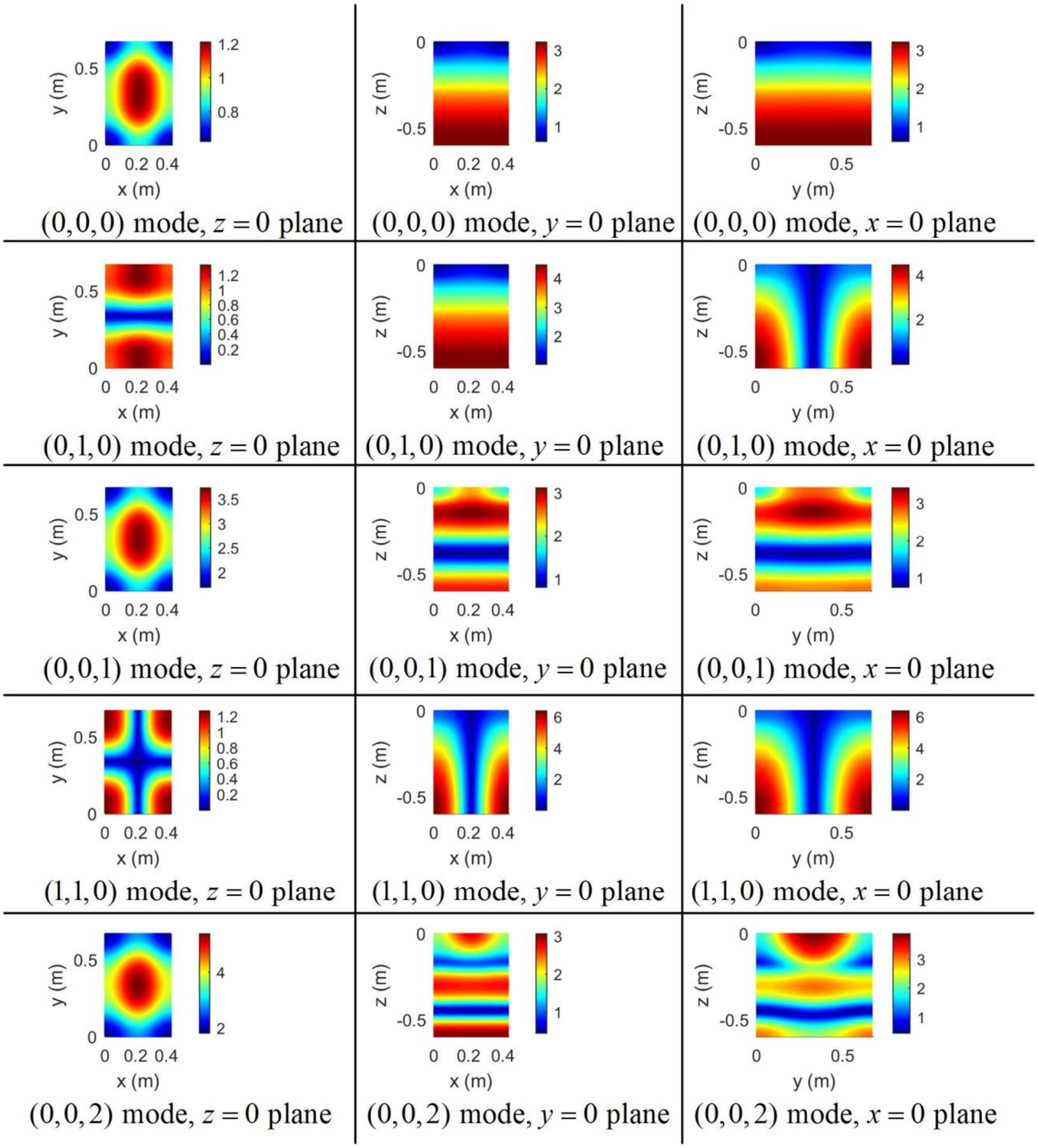

FIG. 2. The modulus of $\Psi_{\mu,\nu,\xi}(\boldsymbol{x})$ (in Pa) for the (0,0,0), (0,1,0), (0,0,1), (1,1,0), and (0.0,2) modes, when the source frequency $f = 500$ Hz.

Figure 3 presents the amplitude of each frequency-dependent eigenmode upon the monopole source with strength $q_s = jk\rho_0 c_0 q_0$ ($q_0 = 10^{-4}$ m$^3$/s and $f = 500$ Hz), where one can see that $|c_{\mu,\nu,\xi}|$ decays rapidly as the order of the mode grows. Here, $|c_{\mu,\nu,\xi}|$ takes the maximum value at the (1,1,0) mode, at which the eigenfrequency takes a value of $484.95 + 1.83j$ Hz. The number of frequency-dependent eigenmodes needed for calculation in Fig. 4 are then examined. This indicates that less than fifteen eigenmodes are required for the sound pressure to converge at the evaluation points inside the cavity, while less than twenty eigenmodes are required for the evaluation points outside the cavity. This result is quite reasonable considering the resonant eigenmode is the $6^{th}$ one.

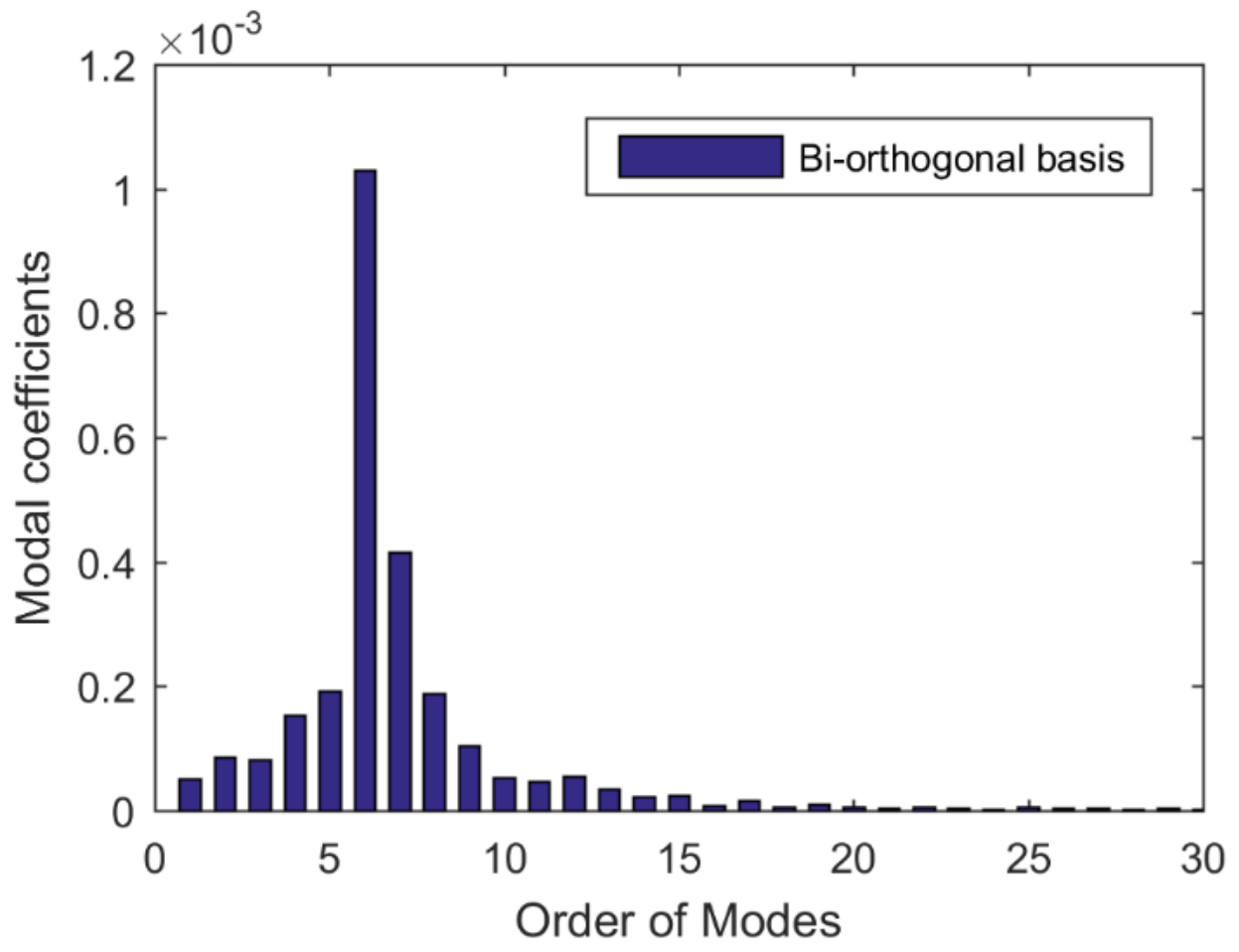

FIG. 3. The amplitudes of modal coefficients $|c_{\mu,\nu,\xi}|$ versus the order of the modes, when the source frequency is 500 Hz.

The performance of the proposed method is verified by calculating the sound pressure at field points inside and outside the cavity for multiple frequencies below 500 Hz, where the calculation is implemented utilizing the first twenty frequency-dependent eigenmodes. The reference result is obtained using the commercial finite-element software COMSOL, where PMLs are used to model the semi-infinite space above the baffle. Note that only frequencies above 30 Hz are treated in COMSOL, as at very low frequencies, the PMLs needed for calculation become very thick in order to prevent spurious wave reflection. The source strength is taken as $q_s = 4\pi \times 10^{-4}$ kg/s$^2$ for all frequencies. Figure 5 plots and compares the results obtained by both methods, in which the excellent agreement verifies the efficacy and accuracy of the proposed method.

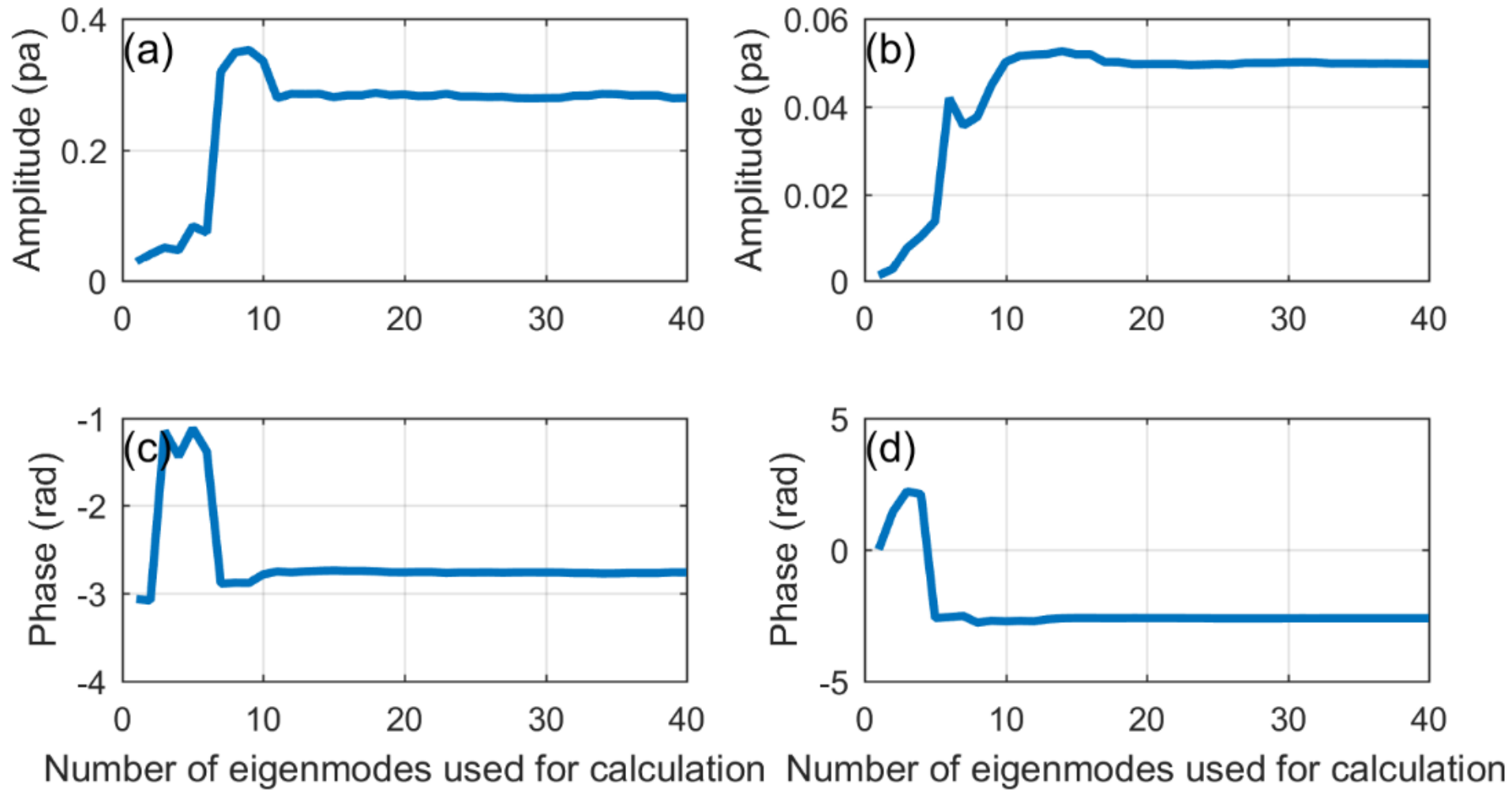

FIG. 4. The amplitude and phase of the sound pressure as a function of the number of eigenmodes used for calculation. The sound source is located at (0.1, 0.1, $L_z$+0.1) m with $f$ = 500 Hz, $q_s = 10^{-4}$ kg/s$^2$. The evaluation points are located, respectively, at location (0.2, 0.3, $L_z$+0.4) m in the cavity (subfigures (a) and (c)) and at location (1.3, 1.4, $-L_z$+1.5) m outside the cavity (subfigures (b) and (d)).

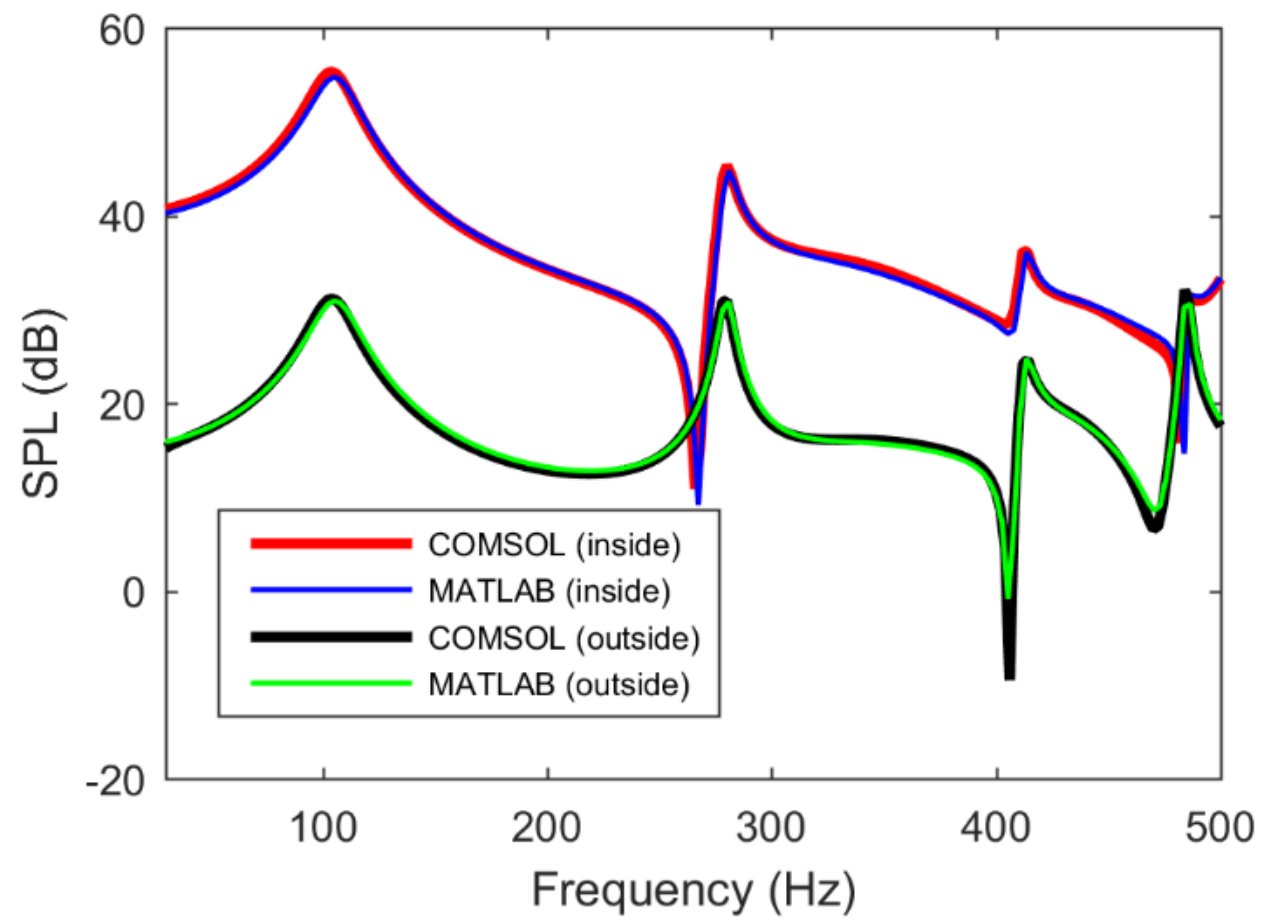

FIG. 5. Comparison of the sound field obtained by the analytical model (marked as MATLAB) and finite-element simulation (marked as COMSOL) when the excitation point source is located at (0.1, 0.1, 0.1) m and $q_s = 4\pi \times 10^{-4}$ kg/s$^2$, for sound pressure levels evaluated at (0.2, 0.3, 0.4) m in the cavity and (1.3, 1.4, 1.5) m outside the cavity.

## III. UNBAFFLED OPEN CAVITY

Analytically, the unbaffled open cavity is much more complicated than the baffled one, where the difficulty lies in the expression of the sound field outside the cavity. However, within the framework of the proposed method, most of the procedures in Sec. II for a baffled cavity can be extended directly to the unbaffled open cavity, except that the basis set for the external space is obtained utilizing numerical tools. This section will be devoted to the forced response of the unbaffled open cavity to a source placed within the cavity.

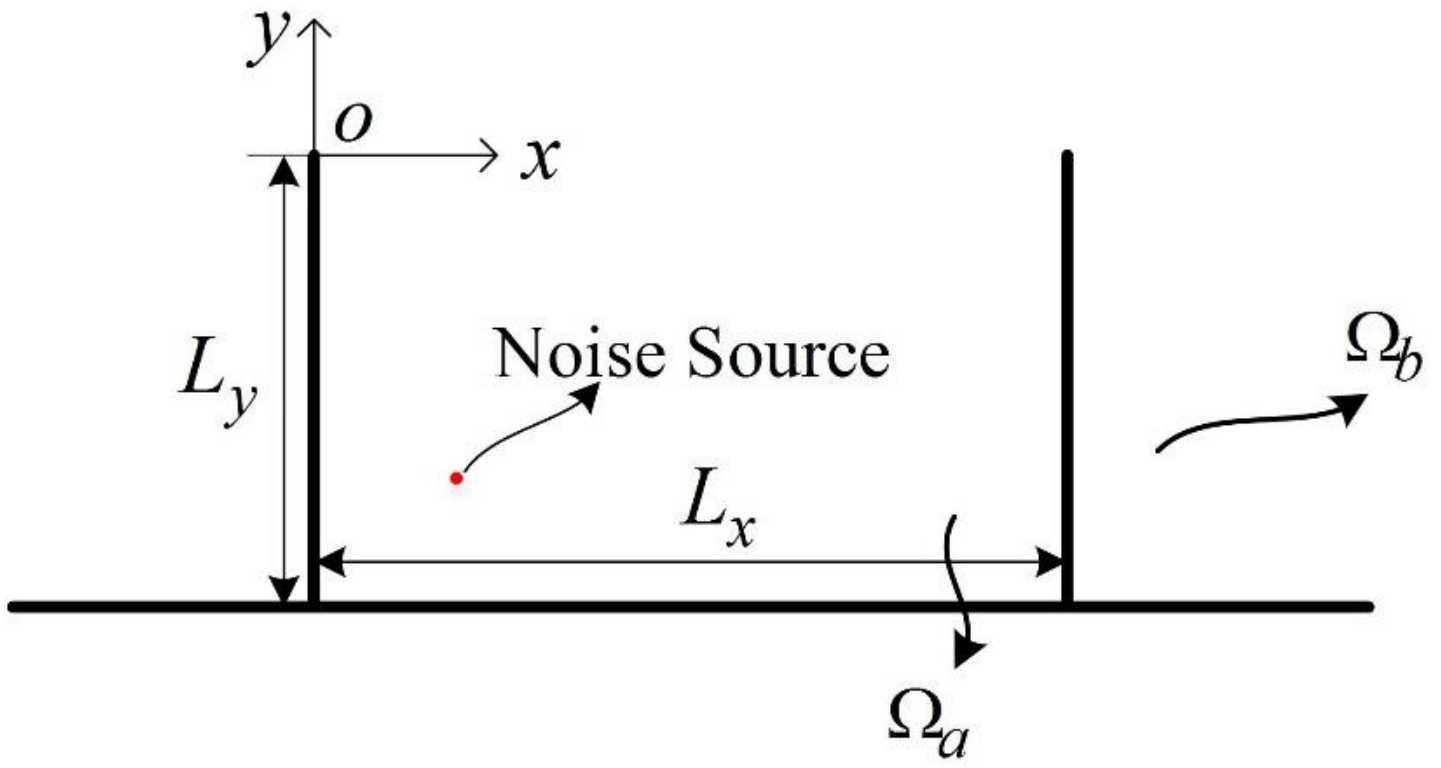

FIG. 6. A 2D unbaffled cavity.

For the convenience of discussion, as well as of visualization, a 2D unbaffled open cavity is considered, as depicted in Fig. 6. In analogy to the solution for the baffled cavity, the basis functions taken for $\Omega_a$ are the closed-cavity modes, analytically written as:

$$\phi_{\mu,\nu}(\boldsymbol{x}) = \psi_\mu(x)\psi_\nu(y). \tag{29}$$

The basis functions for $\Omega_b$ are found by using finite-element analysis (FEA). As shown in Fig. 7, the space $\Omega_b$ is bounded by PMLs and exerting the normal velocity distribution $v_n(x) = \psi_n(x)$ at the opening gives the $n^{\text{th}}$ basis function $\varphi_n(\boldsymbol{x})$ for the expansion of $p_b(\boldsymbol{x})$. Figure 8 presents the modulus values of the first three basis functions in $\{\varphi_n(\boldsymbol{x})\}$ at the source frequency of 300 Hz. The 2D open cavity considered in the numerical computation has a width ($L_x$) of 0.763 m and a height ($L_y$) of 0.531 m. COMSOL is used for the computation of $\{\varphi_n(\boldsymbol{x})\}$.

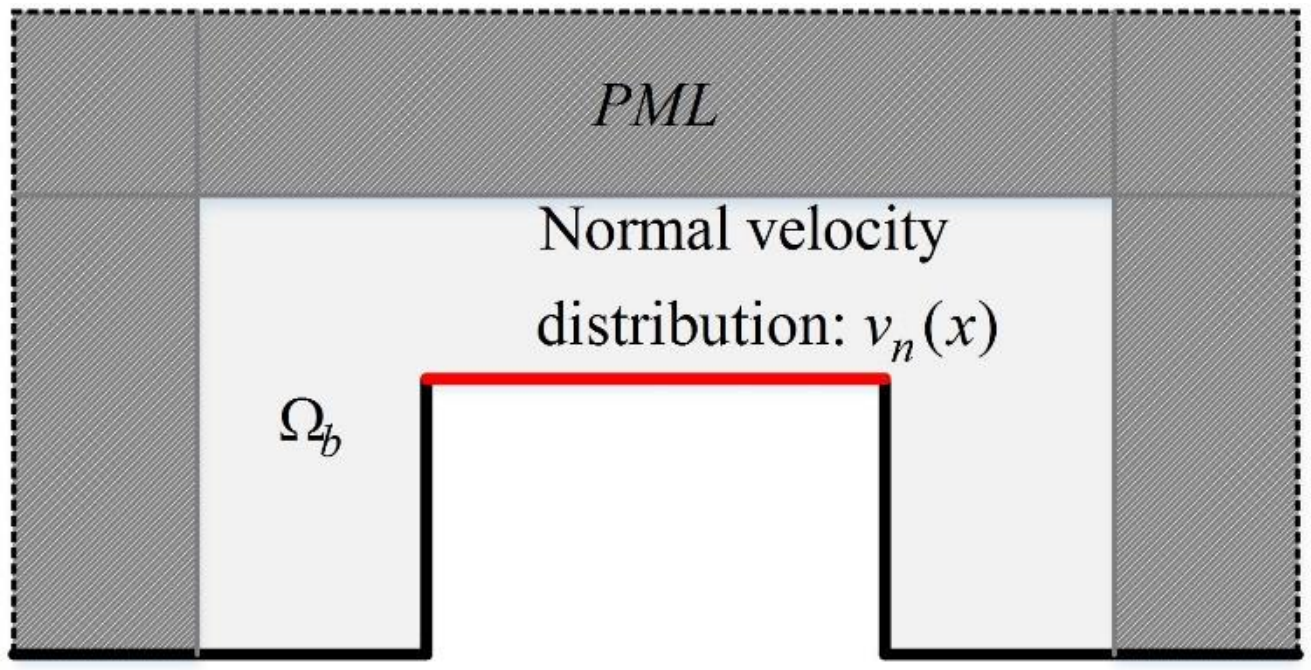

FIG. 7. The calculation of the basis function $\psi_n(\boldsymbol{x})$ for $\Omega_b$: (a) $|\varphi_0(\boldsymbol{x})|$, (b) $|\varphi_1(\boldsymbol{x})|$, and (c) $|\varphi_2(\boldsymbol{x})|$.

Once the basis functions are obtained for the cavity and external areas, the solution is almost the same as described in Sec. II, giving the modal expansion of the sound field as:

$$p(\boldsymbol{x}) = \begin{pmatrix} p_a(\boldsymbol{x}) \\ p_b(\boldsymbol{x}) \end{pmatrix} = \sum_{\mu=0}^{\infty} \sum_{\nu=0}^{\infty} c_{\mu,\nu} \begin{pmatrix} \Phi_{\mu,\nu}(\boldsymbol{x}) \\ \Psi_{\mu,\nu}(\boldsymbol{x}) \end{pmatrix}, \tag{30}$$

where $c_{\mu,\nu}$, $\Phi_{\mu,\nu}(\boldsymbol{x})$, and $\Psi_{\mu,\nu}(\boldsymbol{x})$ take similar forms as their counterparts in the case of the baffled open cavity. Thanks to the utilization of FEA for $\{\varphi_n(\boldsymbol{x})\}$ and the convenience in visualization of a 2D solution, the modulus values of the first six frequency-dependent eigenmodes (including both $\Psi_{\mu,\nu}(\boldsymbol{x})$ and $\Phi_{\mu,\nu}(\boldsymbol{x})$ ) at source frequency $f = 300$ Hz are presented in Fig. 9. Figure 10, on the other hand, provides a comparison of the reference sound field via FEA and the predicted sound pressure distribution based on the first twenty frequency-dependent eigenmodes at $f = 300$ Hz using MATLAB. Twenty closed-cavity modes and five external basis functions are used in the computation of the frequency-dependent eigensolutions; the point source is placed randomly at $(0.2,\ 0.1 - L_y)$ m. Excellent agreement between the results of the proposed method and the reference method is observed. Finally, the performance of the proposed method is examined for multiple frequencies below 500 Hz, still using the first twenty frequency-dependent eigenmodes. Evaluation points are randomly chosen to be at $(0.7,\ 0.3 - L_y)$ m inside the cavity and $(1,\ 0.9 - L_y)$ m outside the cavity. The numerical result is displayed in Fig. 11, which is as anticipated, showing good agreement at frequencies below 500 Hz.

## IV. DISCUSSIONS AND REMARKS

The proposed method can be straightforwardly extended to any general open cavity, like that shown in Fig. 12. Owing to the regular shapes of the cavities considered in Secs. II and III, the basis functions for the space occupied by the cavity, *i.e.*, $\phi_{\mu,\nu,\xi}(\boldsymbol{x})$, were given analytically. In practice, however, $\phi_{\mu,\nu,\xi}(\boldsymbol{x})$ (may use different sub-indeces as the separation of coordiates may not always be possible) can be obtained by solving the closed-cavity modes of $\Omega_a$ through mature numerical (*e.g.*, finite-element method, boundary-element method) eigen solvers, while the basis functions for the external space ($\varphi_{m,n}(\boldsymbol{x})$ in Sec. II and $\varphi_n(\boldsymbol{x})$ in Sec. III) can be obtained following the procedure presented in Sec. III. For the present paper, all boundaries of the open cavity are assumed to be rigid, but non-rigid boundaries can be taken into account as well by referring to the treatment in Ref. 11. This has potential applications in a variety of problems, *e.g.*, sound barriers, noise radiation from enclosures, *etc*. Compared to traditional numerical methods, the proposed method is semi-analytical and may reduce computational load in problems such as the optimization of secondary sources in the active noise control of enclosures. Furthermore, it may provide some physical understanding of acoustic coupling between cavities and an external space.

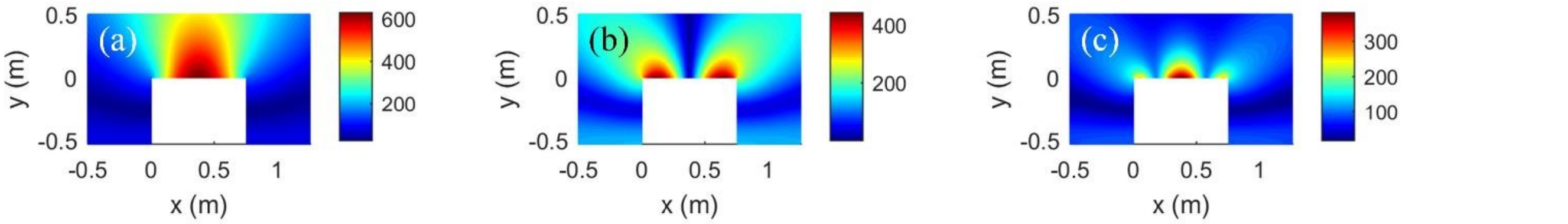

FIG. 8. The modulus values of the first three basis functions of $p_b(\boldsymbol{x})$, at source frequency $f = 300$ Hz.

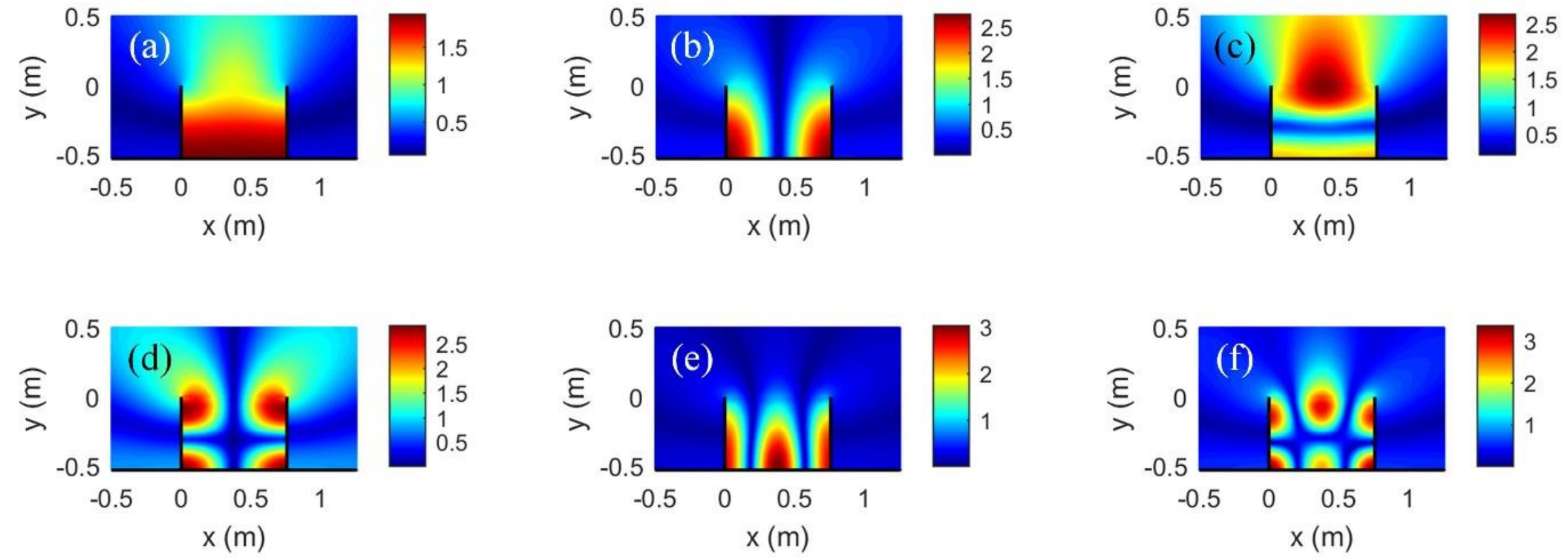

FIG. 9. The first six frequency-dependent eigenmodes of the unbaffled cavity, at source frequency $f = 300$ Hz: (a) $|\Psi_{0,0}(\boldsymbol{x})|$ and $|\Phi_{0,0}(\boldsymbol{x})|$, (b) $|\Psi_{1,0}(\boldsymbol{x})|$ and $|\Phi_{1,0}(\boldsymbol{x})|$, (c) $|\Psi_{0,1}(\boldsymbol{x})|$ and $|\Phi_{0,1}(\boldsymbol{x})|$, (d) $|\Psi_{1,1}(\boldsymbol{x})|$ and $|\Phi_{1,1}(\boldsymbol{x})|$, (e) $|\Psi_{2,0}(\boldsymbol{x})|$ and $|\Phi_{2,0}(\boldsymbol{x})|$, and (f) $|\Psi_{2,1}(\boldsymbol{x})$ and $|\Phi_{2,1}(\boldsymbol{x})|$.

The assumption that a sound source is placed within a cavity greatly simplifies the problem by enabling the usage of an incomplete basis function set for sound expansion outside the cavity. However, it also limits the proposed method from solving a general open-cavity problem where sound sources are placed both inside and outside the cavity, as the incomplete basis functions cannot fully represent the external sound field if it has a sound source within. The modal solution to such a general open-cavity problem still remains an open question, and may be solved within the framework of coupled mode theory and frequency-dependent eigenmodes, if a proper complete basis function set is chosen for the external space.

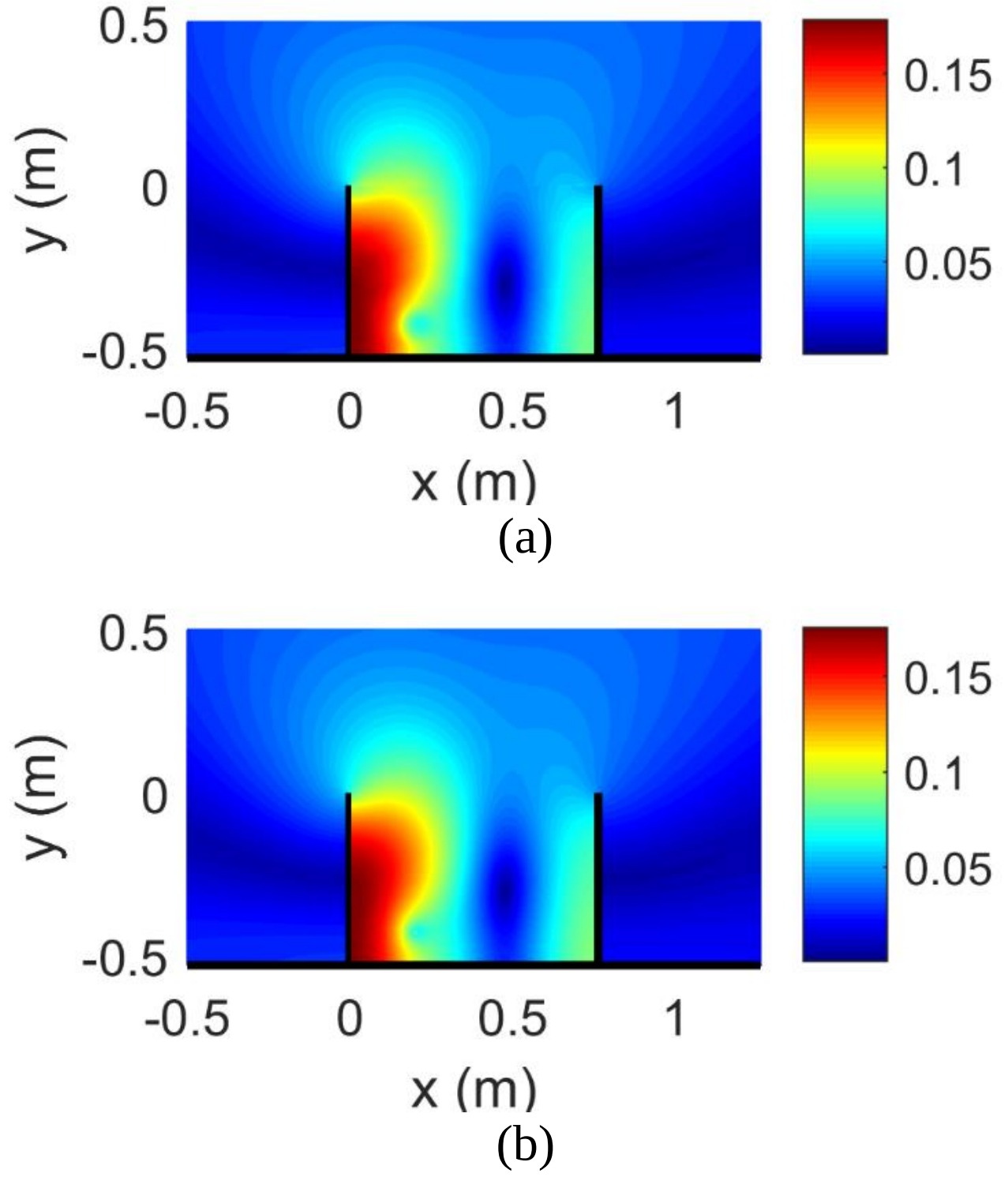

FIG. 10. The modulus of sound pressure distribution of the unbaffled open cavity, $|p(\boldsymbol{x})|$, at source frequency $f = 300$ Hz: (a) reconstructed field and (b) reference sound pressure field obtained using COMSOL.

Finally, a remark is made on an eigenvalue problem similar to Eq. (19):

$$\boldsymbol{D}(\tilde{K}_{\mu,\nu,\xi}^2)\tilde{\boldsymbol{g}}_{\mu,\nu,\xi} = \tilde{K}_{\mu,\nu,\xi}^2\tilde{\boldsymbol{g}}_{\mu,\nu,\xi}, \tag{31}$$

where the matrix $\boldsymbol{D}$ is eigenvalue-dependent rather than $k$-dependent. It corresponds to the natural vibration of the system in the absence of a noise source, which is the *acoustic resonance* commonly encountered in the literature, *e.g.*, Ref. 5. Equation (31) can be solved directly[14] and is equivalent to the solution found using a finite-element eigen solver.[18] As mentioned in the Introduction, the eigensolutions to Eq. (31) are non-orthogonal and may not be complete for a modal representation of the forced response of the system. The reader may refer to Ref. 10 for more discussion on this issue.

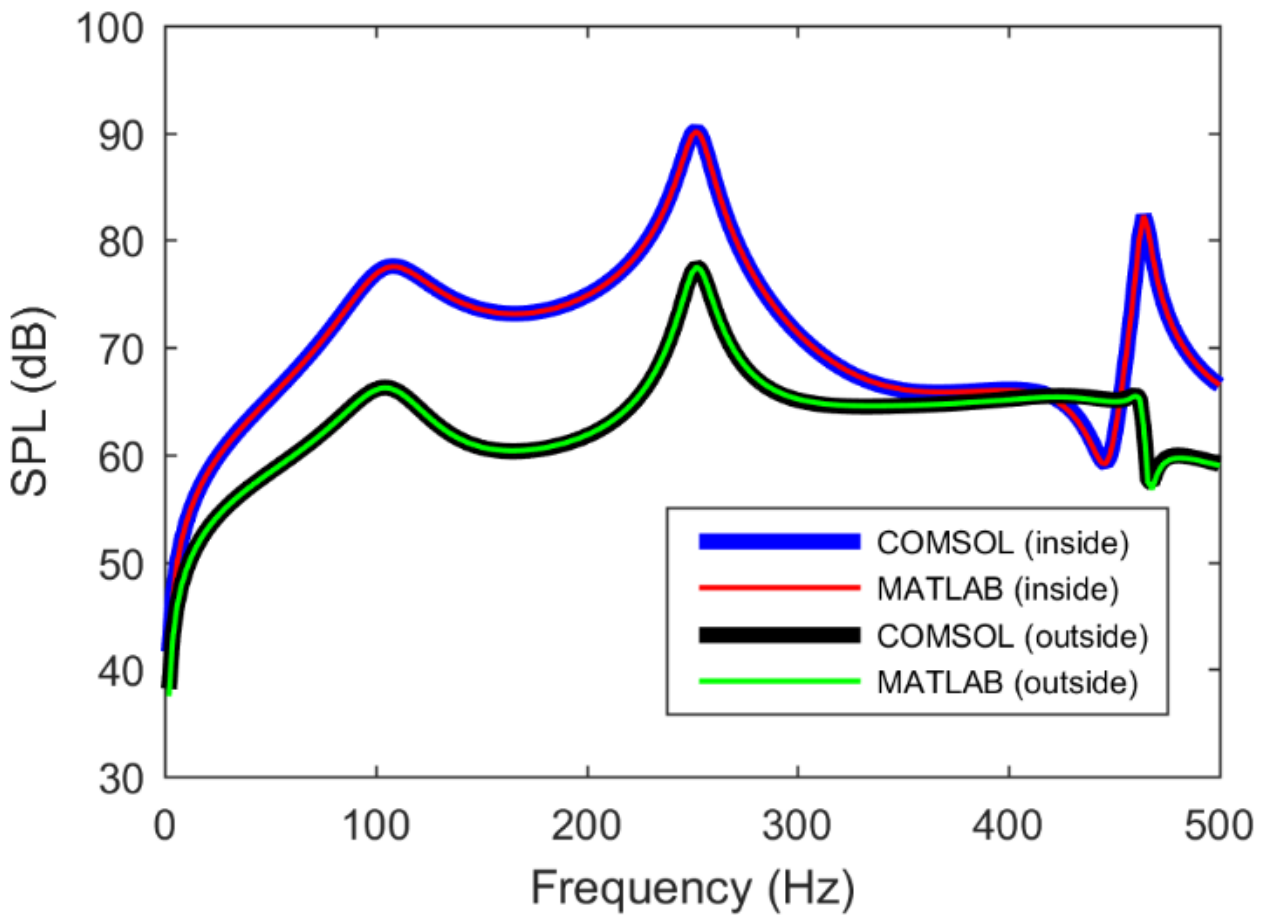

FIG. 11. Comparison of sound fields obtained by the analytical model (marked as MATLAB) and the finite-element simulation (marked as COMSOL) when the excitation point source is located at $(0.2,\ 0.1 - L_y)$ m and $q_0 = 10^{-4}$ m$^2$/s: sound pressure level at $(0.7,\ 0.3 - L_y)$ m in the cavity and $(1,\ 0.9 - L_y)$ m outside the cavity.

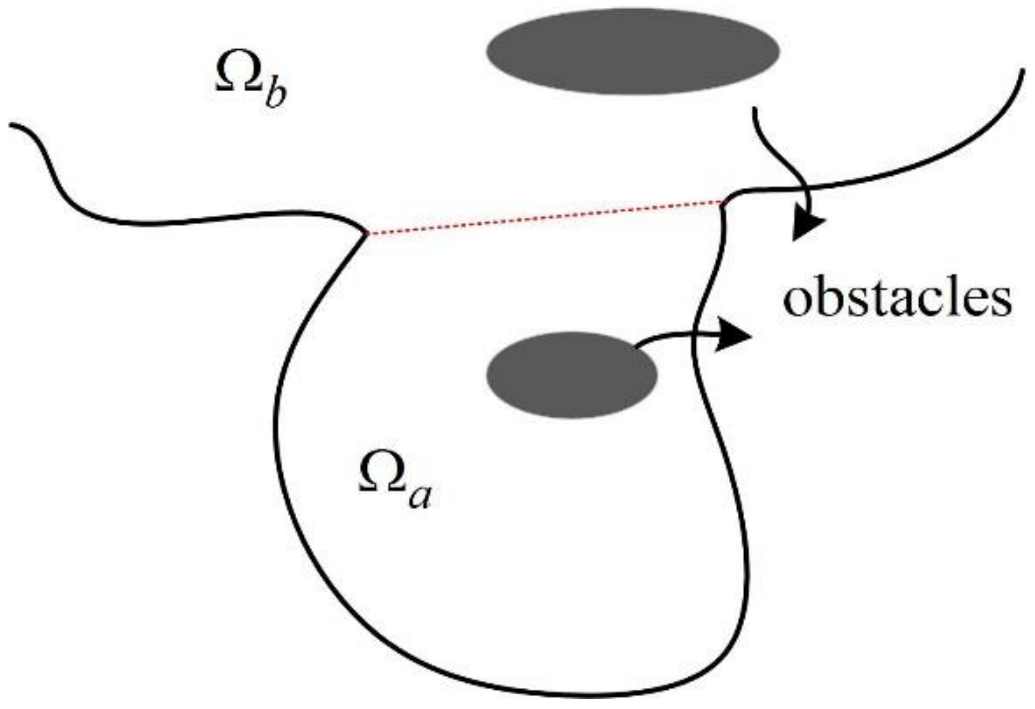

FIG. 12. A general open cavity.

## V. CONCLUSIONS

The forced acoustical response of a baffled rectangular open cavity with a sound source within the cavity was formulated as the superposition of the bi-orthogonal frequency-dependent eigenmodes. The effectiveness of the modal expansion was demonstrated numerically by showing that an accurate sound pressure prediction can be obtained using only a few frequency-dependent eigenmodes for expansion of evaluation points either inside or outside the cavity. The proposed modal expansion was also extended to unbaffled open cavities where a finite-element method was employed for the computation of the basis functions of the external field.

Discussions were then made on the advantages as well as limitations of the proposed method. A remark was finally given to distinguish the frequency-dependent eigenmodes from frequency-independent eigenmodes.

## Acknowledgements

The authors wish to thank Shuping Wang from Nanjing University for discussion on the calculation of radiation impedance. The financial support from the Australian Research Council (ARC LP) is gratefully acknowledged. The first author is also grateful for the sponsorship from the Chinese Scholarship Council.